\documentclass[11pt]{article}
\pdfoutput=1
\usepackage{dcolumn}% Align table columns on decimal point
\usepackage{bm}% bold math
\usepackage{graphicx}% Include figure files
\usepackage{amssymb,amsmath}
\usepackage{multirow}
\usepackage{cite,color,url}
\usepackage[colorlinks=true
,urlcolor=blue
,anchorcolor=blue
,citecolor=blue
,filecolor=blue
,linkcolor=blue
,menucolor=blue
,linktocpage=true
,pdfproducer=medialab
,pdfa=true
]{hyperref}

\usepackage{slashed}
\usepackage{epsfig,psfrag,rotating,soul}
\usepackage{rotfloat}
\evensidemargin \oddsidemargin
\allowdisplaybreaks

\let\OLDthebibliography\thebibliography
\renewcommand\thebibliography[1]{
  \OLDthebibliography{#1}
  \setlength{\parskip}{0pt}
  \setlength{\itemsep}{0pt plus 0.3ex}
}

\begin{document}
\thispagestyle{empty}

\def\thefootnote{\fnsymbol{footnote}}

\begin{flushright}
  CIFFU-17-06\\
\end{flushright}

\vspace*{1cm}

\begin{center}

\begin{Large}
\textbf{\textsc{Search strategies for pair production of heavy Higgs bosons \\[.25em] decaying invisibly at the LHC}}
\end{Large}

\vspace{1cm}

{\sc
E.~Arganda$^{1}$%
\footnote{{\tt \href{mailto:ernesto.arganda@fisica.unlp.edu.ar}{ernesto.arganda@fisica.unlp.edu.ar}}}%
, J.L.~Diaz-Cruz$^{2,3}$%
\footnote{{\tt \href{mailto:jldiaz@fcfm.buap.mx}{jldiaz@fcfm.buap.mx}}}%
, N.~Mileo$^{1}$%
\footnote{{\tt \href{mailto:mileo@fisica.unlp.edu.ar}{mileo@fisica.unlp.edu.ar}}}%
, R.A.~Morales$^{1}$%
\footnote{{\tt \href{mailto:roberto.morales@fisica.unlp.edu.ar}{roberto.morales@fisica.unlp.edu.ar}}}%
, A.~Szynkman$^{1}$%
\footnote{{\tt \href{mailto:szynkman@fisica.unlp.edu.ar}{szynkman@fisica.unlp.edu.ar}}}%
}

\vspace*{.7cm}

{\sl
$^1$IFLP, CONICET - Dpto. de F\'{\i}sica, Universidad Nacional de La Plata, \\ 
C.C. 67, 1900 La Plata, Argentina

\vspace*{0.1cm}

$^2$Centro Internacional de F\'{\i}sica Fundamental, Benem\'erita Universidad Aut\'onoma de Puebla \\

\vspace*{0.1cm}

$^3$Facultad de Ciencias F\'{\i}sico-Matem\'aticas, \\
 Benem\'erita Universidad Aut\'onoma de Puebla, Puebla, M\'exico
}

\end{center}

\vspace{0.1cm}

\begin{abstract}
\noindent

The search for heavy Higgs bosons at the LHC represents an intense experimental program, carried out by the ATLAS and CMS collaborations, which includes the hunt for invisible Higgs decays and dark matter candidates. No significant deviations from the SM backgrounds have been observed in any of these searches, imposing significant constraints on the parameter space of different new physics models with an extended Higgs sector. Here we discuss an alternative search strategy for heavy Higgs bosons decaying invisibly at the LHC, focusing on the pair production of a heavy scalar $H$ together with a pseudoscalar $A$, through the production mode $q \bar q \to Z^\ast \to HA$. We identify as the most promising signal the final state made up of $4b + E_T^\text{miss}$, coming from the heavy scalar decay mode $H \to hh \to b \bar b b \bar b$, with $h$ being the discovered SM-like Higgs boson with $m_h=125$ GeV, together with the invisible channel of the pseudoscalar. We work within the context of simplified MSSM scenarios that contain quite heavy sfermions of most types with ${\cal O}(10)$ TeV masses, while the stops are heavy enough to reproduce the 125 GeV mass for the lightest SM-like Higgs boson. By contrast, the gauginos/higgsinos and the heavy MSSM Higgs bosons have masses near the EW scale. Our search strategies, for a LHC center-of-mass energy of $\sqrt{s} =$ 14 TeV, allow us to obtain statistical significances of the signal over the SM backgrounds with values up to $\sim$ 1.6$\sigma$ and $\sim$ 3$\sigma$, for total integrated luminosities of 300 fb$^{-1}$ and 1000 fb$^{-1}$, respectively.

\end{abstract}

\def\thefootnote{\arabic{footnote}}
\setcounter{page}{0}
\setcounter{footnote}{0}

\newpage

\section{Introduction}
\label{intro}
 
The  discovery at the LHC of a SM-like Higgs boson~\cite{Aad:2012tfa,Chatrchyan:2012xdj}, with the most recent measurement 
of its mass set at $m_{h_\text{SM}} =$ 125.09 $\pm$ 0.21 (stat.) $\pm$ 0.11 (syst.) GeV~\cite{Aad:2015zhl}, 
has been a major confirmation of the standard model of particle physics (SM). The experiments reveal that the Higgs boson mass value agrees quite well with the range preferred by the analysis of electroweak precision tests~\cite{Flacher:2008zq,Erler:2010wa}, and the spin, parity, and some of its couplings to the SM particles do not show deviations from the standard expectations~\cite{HiggsCouplingsCombined}. Further studies of the Higgs couplings are required in order to test more precisely its SM nature~\cite{Espinosa:2012im,Giardino:2012ww}, or to find evidence of physics beyond the SM (BSM). In fact, the LHC has already provided important bounds on the scale of new physics~\cite{Ellis:2010wx}.

A clear evidence of BSM physics would be doubtlessly the existence of new Higgs bosons, as predicted in some of the simplest extensions of the SM: the general two-Higgs-doublet models (2HDM) (for a review, see, e.g.,~\cite{Branco:2011iw}) or the Higgs sector of the minimal supersymmetric standard model (MSSM) (for reviews, see, e.g.,~\cite{Nilles:1983ge,Haber:1984rc}). Within these classes of models, the extended Higgs sector includes five Higgs physical states: two CP-even bosons (one light scalar $h$ and one heavy scalar $H$), as well as one CP-odd boson (the pseudoscalar $A$) and a charged Higgs bosons pair ($H^\pm$). In this context, the lightest Higgs boson $h$ is usually identified with the discovered SM-like Higgs boson of 125 GeV and there is an intense experimental program, performed by the ATLAS and CMS collaborations, in order to search for the remaining heavy Higgs bosons at the LHC, that depends on the different Higgs decay channels considered. The heavy neutral Higgs bosons $H$ and $A$ are being looked for through their decay modes into vector and/or SM-like Higgs bosons ($\gamma\gamma$~\cite{Aad:2014ioa,Khachatryan:2014jya,Khachatryan:2015qba,Aaboud:2016tru,Aaboud:2017yyg}, $Z\gamma$~\cite{Aaboud:2016trl,Aaboud:2017uhw}, $ZZ$~\cite{Aad:2015kna,Aaboud:2016okv}, $W^+W^-$~\cite{Aaboud:2016okv,Aaboud:2017gsl}, $Zh$~\cite{Aad:2015wra,Khachatryan:2015lba,Khachatryan:2015tha,Aaboud:2017ahz}, and $hh$~\cite{Khachatryan:2015tha,Aad:2014yja,Khachatryan:2015yea,Aad:2015uka,Khachatryan:2015wka,Aad:2015xja,Aaboud:2016xco,Sirunyan:2017tqo,Sirunyan:2017djm,Sirunyan:2017guj}), into charged-lepton pairs ($\mu^+\mu^-$~\cite{CMS:2015ooa} and $\tau^+\tau^-$~\cite{Chatrchyan:2011nx,Aad:2011rv,Chatrchyan:2012vp,Aad:2012cfr,Khachatryan:2014wca,Aad:2014vgg,Aaboud:2016cre,Aaboud:2017sjh}), and into heavy-flavor quarks ($b \bar b$~\cite{Khachatryan:2015tra} and $t \bar t$~\cite{Aaboud:2017hnm}). Meanwhile, the charged Higgs bosons are searched for in the $H^\pm \to W^\pm Z$~\cite{Aad:2015nfa,Sirunyan:2017sbn}, $H^\pm \to \tau^\pm\nu$~\cite{Aad:2012tj,Aad:2014kga,Khachatryan:2015qxa,Aaboud:2016dig}, and $H^\pm \to tb$~\cite{Khachatryan:2015qxa,Aad:2015typ} decay channels. No significant deviations from the SM backgrounds have been found in any of these searches, imposing important constraints on the parameter space of simplified models of extended Higgs sectors, as the so-called $h$MSSM~\cite{Djouadi:2013vqa,Djouadi:2013uqa,Djouadi:2013lra,Djouadi:2015jea,Chalons:2017wnz}, but without taking the possibility of invisible Higgs decays into account.

On the other hand, the presence of invisible Higgs decays, which would be another clear signal of new physics, is very well-motivated and predicted in many extensions of the SM, such as the MSSM (for a review, see, e.g.,~\cite{Curtin:2013fra}). The searches for these exotic decays at the LHC are centered on the production of an invisibly decaying Higgs boson via gluon fusion~\cite{Khachatryan:2016whc}, vector boson fusion~\cite{Chatrchyan:2014tja,Aad:2015txa,Aad:2015pla,Khachatryan:2016whc}, and in association with a vector boson~\cite{Aad:2014iia,Chatrchyan:2014tja,Aad:2015uga,Aad:2015pla,Khachatryan:2016whc,Aaboud:2017bja}. These searches are also carried out considering the production of dark matter candidates in association with a vector boson~\cite{Aaboud:2017bja,Khachatryan:2016mdm}, with a Higgs boson~\cite{Aad:2015yga,Aad:2015dva,Aaboud:2016obm,Sirunyan:2017hnk,Aaboud:2017uak,Aaboud:2017yqz}, or with heavy-flavor quarks~\cite{Sirunyan:2017xgm,Aaboud:2017rzf}. As it happens with the searches for heavy Higgs bosons, no significant excesses have been observed over the SM backgrounds in these invisible decay searches and limits are placed on the parameter spaces of the different models, production cross sections, and invisible branching ratios.

It is worth to stress that none of the searches listed above considers the heavy Higgs boson pair production. With the aim to explore this production mechanism, we analyze in a jointed framework the invisible heavy Higgs decays together with the decays into SM particles as final products. In particular, this combination represents a probe of the coupling between two heavy Higgs bosons and the dark matter particles taking part of the hard interaction. Remarkably, besides the gluon fusion plus jets channel, the heavy Higgs-pair production is unique to study invisible decays of the pseudoscalar. Without the vector boson fusion and bremsstrahlung modes, which are absent for the pseudoscalar, the other possible production mechanisms are the associated with a light Higgs and with a pair of quarks. However, the former is dynamically suppressed within the MSSM whereas the latter has a cross section at the edge of accessibility at HL-LHC. We propose then an alternative search strategy for the heavy neutral Higgs bosons, by means of the production of a pair of Higgs bosons $H+A$ through the tree-level mode $q \bar q \to Z^\ast \to HA$, and taking the possibility of invisible Higgs decays into account. In order to make quantitative statements, we work within a particular MSSM scenario, so-called Slim SUSY~\cite{Arganda:2012qp,Arganda:2013ve}, but the conclusions are general for any given scenario with a similar mass spectrum. In the Slim SUSY scenario the only new particles at the electroweak (EW) scale are the heavy Higgs bosons and the charginos and/or the neutralinos. Within the context of Slim SUSY, the $R$ parity is conserved and therefore the lightest supersymmetric particle (LSP), which is assumed to be the lightest neutralino $\chi$, is stable, which allows for invisible Higgs decays $H, A \to \chi \chi$ if $M_{H,A} > 2M_\chi$. In the low-tan$\beta$ regime, these invisible decay channels can be sizeable, and even the dominant decay mode of the pseudoscalar, as we will show below. This low-tan$\beta$ regime~\cite{Bagnaschi:2015hka}, recently noticed in~\cite{Arganda:2012qp,Arganda:2013ve} and highlighted in~\cite{Djouadi:2013vqa,Djouadi:2013uqa,Djouadi:2013lra,Djouadi:2015jea,Chalons:2017wnz,Lee:2015uza,Craig:2016ygr}, gives rise to a rich Higgs phenomenology since the decay rates of the heavy Higgs bosons are not dominated by the decay channels into bottom-quark and $\tau$-lepton pairs, as occurs for moderate and large values of $\tan\beta$. This situation opens the possibility of heavy Higgs boson searches through interesting channels, as $H \to W^+ W^-$, $H \to ZZ$, $H \to hh$, $A \to hZ$, the decay modes into top-quark pairs $H, A \to t \bar t$, and the invisible decays $H, A \to \chi\chi$.

The paper is organized as follows: in Section~\ref{th-framework} we present the MSSM scenarios in the low-$\tan\beta$ regime with invisible Higgs decays that give rise to the mentioned rich Higgs phenomenology. Section~\ref{HAproduction} is devoted to the heavy Higgs-pair production modes at the LHC, the description of the different potential final states, and the confrontation with the current LHC searches. In Section~\ref{strategy} a dedicated search strategy for the promising channel $pp \to HA \to 4b+E_T^{\text{miss}}$ is performed, showing the expected significance of the signal at the LHC and the future prospects at the HL-LHC. Finally, perspectives and conclusions are presented in Section~\ref{conclusions}.

\section{MSSM scenarios in the low-\texorpdfstring{$\tan\beta$}{tanb} regime with invisible Higgs decays}
\label{th-framework}

The rich phenomenology that arises in the MSSM Higgs sector, when the value of $\tan\beta$ is low, was recently remarked in~\cite{Arganda:2012qp,Arganda:2013ve} and emphasized in~\cite{Djouadi:2013vqa,Djouadi:2013uqa,Djouadi:2013lra,Djouadi:2015jea,Chalons:2017wnz,Lee:2015uza,Craig:2016ygr}. In the low-$\tan\beta$ regime, the Higgs phenomenology is significantly enriched due to the suppression of the heavy neutral MSSM Higgs bosons decays into down-type fermions, which are proportional to $\tan\beta$. In such a case, these heavy Higgs bosons will have sizeable branching ratios for a variety of interesting decay modes, depending on their masses one could have: $H \to$ $W^+W^-$, $ZZ$, $hh$, $t \bar t$, and $A \to$ $hZ$, $t \bar t$. Another possibility, which was not considered in~\cite{Djouadi:2013vqa,Djouadi:2013uqa,Djouadi:2013lra,Djouadi:2015jea,Chalons:2017wnz,Lee:2015uza,Craig:2016ygr}, is that the heavy scalar $H$ and the pseudoscalar $A$ decay invisibly into a pair of lightest neutralinos, $H, A \to \chi\chi$ (if $M_\chi < M_{H,A}/2$), assumed to be the LSP and consequently stable. These invisible decays can also have important branching ratios in the low-$\tan\beta$ regime and they could even be the dominant decay channel for the pseudoscalar.

In order to study this rich Higgs phenomenology, in the low-$\tan\beta$ regime with invisible Higgs decays, we work within the context of Slim SUSY scenarios~\cite{Arganda:2012qp,Arganda:2013ve}, defined with the following assumptions:
\begin{enumerate}
\item It contains heavy stops, with large enough masses to account for the Higgs mass value ($m_{h_\text{SM}} \simeq$ 125 GeV) through radiative corrections.
\item Heavy masses of the first and second generation sfermions to solve the SUSY and CP flavor problems or at least to ameliorate them~\cite{ArkaniHamed:1997ab,DiazCruz:2005qz}.
\item A neutralino sector with an LSP $m_{\chi} =$ ${\cal O}$(100 GeV) is chosen as the dark matter candidate~\cite{Baer:2011ab}.
\item The full Higgs sector has masses near the EW scale.
\end{enumerate}

This class of simplified MSSM scenarios (which we can label as Slim Pheno) can be described with only four parameters at the EW scale: $\tan\beta$, the pseudoscalar mass $M_A$, the bino mass $M_1$, and the higgsino mass $\mu$\footnote{The other two gaugino masses, the wino mass $M_2$ and the gluino mass $M_3$, are set at the TeV scale, and do not play any role in the phenomenology studied along this work.}. We choose $\mu$ to be close enough to $M_1$ in order to obtain an important bino/higgsino admixture, which allows to have sizable Higgs-neutralino-neutralino couplings~\cite{Djouadi:2001fa}, but large enough to avoid the heavy Higgs decays into a pair of second neutralinos ($\chi_2$) or higgsino-like charginos ($\chi^\pm$). Therefore, the only new particles present at low energies, relevant for the Higgs phenomenology that we are interested in, are the two heavy neutral Higgs bosons ($H$ and $A$) and the LSP neutralino, under the condition $M_\chi < M_{H,A}/2$. All along this work, we fix $\tan\beta =$ 3 as a reference value, and define three different benchmarks accordingly to the value of $M_A$, which generate the following mass spectra, computed with the {\tt SUSY-HIT} package~\cite{Djouadi:2006bz}:
\begin{itemize}
\item {\it Light-mass} scenario: $M_A =$ 200 GeV, $M_H =$ 216 GeV, $M_\chi =$ 60 GeV ($M_1 =$ 75 GeV, \\$\mu =$ 154 GeV).
\item {\it Moderate-mass} scenario: $M_A =$ 300 GeV, $M_H =$ 309 GeV, $M_\chi =$ 129 GeV ($M_1 =$ 150 GeV, $\mu =$ 200 GeV).
\item {\it Heavy-mass} scenario: $M_A =$ 400 GeV, $M_H =$ 406 GeV, $M_\chi =$ 174 GeV ($M_1 =$ 200 GeV, \\$\mu =$ 235 GeV).
\end{itemize}

\begin{table}
\begin{center}
\begin{tabular}{|c|c|c|c|}
\hline
$M_A$ [GeV] & 200 & 300 & 400 \\ \hline \hline
$H(A) \to b \bar b$ & 12\% (33\%) & 13\% (27\%) & 4\% (2\%) \\ \hline
$H(A) \to \tau^+ \tau^-$ & 1\% (4\%) & 2\% (4\%) & 0.6\% (0.3\%) \\ \hline
$H(A) \to t \bar t$ & C (C) & C (C) & 77\% (91\%) \\ \hline
$H (A) \to W^+ W^-$ & 59\% (F) & 26\% (F) & 5\% (F) \\ \hline
$H (A) \to ZZ$ & 23\% (F) & 12\% (F) & 2\% (F) \\ \hline
$H(A) \to hh (hZ)$ & C (C) & 45\% (22\%) & 11\% (2\%) \\ \hline
$H(A) \to \chi\chi$ & 3\% (62\%) & 2\% (46\%) & 0.6\% (4\%) \\ \hline
\end{tabular}
\caption{Branching ratios of the main decay modes of the heavy neutral MSSM Higgs bosons in the low-$\tan\beta$ regime, for the three scenarios {\it light-mass}, {\it moderate-mass}, and {\it heavy-mass} ($\tan\beta =$ 3 and $M_A =$ 200, 300, and 400 GeV, respectively), computed with {\tt SUSY-HIT}~\cite{Djouadi:2006bz}. ``C'' and ``F'' stand for kinematically closed and forbidden, respectively. The values in parentheses correspond to the pseudoscalar Higgs boson $A$ branching ratios.}\label{BRs}
\end{center}
\end{table}

The values of $M_A$ chosen here might be in tension with flavor physics observables as $B \to X_s \gamma$, due to the important contributions to the $b \to s \gamma$ transition coming from the charged Higgs bosons~\cite{Arbey:2017gmh}. However, in the MSSM the different Higgs, chargino, and gluino contributions to $B \to X_s \gamma$ are generically competitive and it is not difficult to obtain cancellation patterns among them~\cite{Altmannshofer:2012ks}. In addition, if the correction factor $\Delta_b$ to the Higgs-bottom Yukawa coupling is positive, the $B \to X_s \gamma$ rate predicted in the MSSM is smaller than in the Type-II 2HDM~\cite{Bechtle:2016kui}, as considered in~\cite{Arbey:2017gmh}. Although a detailed study of the constraints imposed by this class of low energy observables is beyond the scope of this work, we have checked that the predictions for BR($B \to X_s \gamma$) in our benchmarks are allowed at the 2$\sigma$ uncertainty.

We show in Table~\ref{BRs} the values of the branching ratios of the main decay channels of the heavy neutral MSSM Higss bosons, $H$ and $A$, in these three simplified scenarios, calculated also with {\tt SUSY-HIT}~\cite{Djouadi:2006bz}. For $M_A = 200$ GeV, the decay modes $H, A \to t \bar t$, $H \to hh$, and $A \to hZ$ are kinematically closed, and consequently the dominant decay of the heavy scalar $H$ is into $W^+ W^-$ (59\%), followed by the channels into $ZZ$ (23\%) and $b \bar b$ (12\%), while the dominant decay mode of the pseudoscalar is, interestingly, its invisible channel with a branching ratio of 62\%. On the other hand, for $M_A = 300$ GeV, the decay channel $H \to hh$ is open and becomes the dominant one for the $H$ Higgs boson, with a branching ratio of 45\%. Whilst in this kinematic region the decay $A \to hZ$ is also accessible, with a decay rate of 22\%, the invisible $A$ decay mode keeps as the dominant one with a branching ratio of 46\%. Finally, if the threshold for the production of a top-quark pair is open (which happens for $M_A = 400$ GeV), the dominant decay channel of both $H$ and $A$ is into $t \bar t$ (77\% and 91\%, respectively). Notice that in this case the decay $H \to hh$ stays with a relatively significant branching ratio (11\%), and the invisible decay of the pseudoscalar too (4\%).

\section{Heavy Higgs-pair production \texorpdfstring{$H+A$}{HA} at the LHC}
\label{HAproduction}

As discussed in the previous sections, a relevant Higgs production mechanism at the LHC to accomplish our proposal is the Higgs-pair production $H+A$. This process occurs at tree-level and it is mediated by a virtual $Z^{*}$ coming from light-quark annihilation or via gluon fusion at one loop through a box diagram; both have been previously studied in~\cite{Christensen:2012si} for the different pairs of MSSM Higgs bosons. In particular, it is worth to note a significant advantage for the tree-level channel $q \bar q \to Z^\ast \to HA$. Since its amplitude goes as $\sin (\beta-\alpha)$, which is now constrained from the light Higgs search at the LHC to be of order one, there is not a dynamical suppression as in the case of $q \bar q \to Z^\ast \to hA$. In what follows, we will present the corresponding cross sections for the three benchmarks defined in the previous section, discussing the different final states at the LHC in each case. Next, we will probe these three scenarios confronting them with the general searches of new physics at the LHC at 8 TeV and 13 TeV.

\subsection{Cross sections and final states}

\begin{table}
\begin{center}
\begin{tabular}{|c|c|c|c|}
\hline
$M_A$ [GeV] & 200 & 300 & 400 \\ \hline \hline
$\sigma(q \bar q \to HA)$ [fb] & 20.6 & 4.60 & 1.41 \\ \hline
$\sigma(gg \to HA)$ [fb] & 1.33 & 0.08 & 0.01 \\ \hline
\end{tabular}
\caption{Cross sections (in fb) of the heavy Higgs-pair production modes $q \bar q \to HA$ and $gg \to HA$ at the LHC with a center-of-mass energy of $\sqrt{s} =$ 14 TeV, for $M_A =$ 200, 300, and 400 GeV, and $\tan\beta =$ 3, computed at NLO with {\tt HPAIR}~\cite{Spira:1997dg,Spira:2016ztx,Spira:website}.}\label{Xsections}
\end{center}
\end{table}

The cross sections for the heavy Higgs-pair production modes $q \bar q \to HA$ and $gg \to HA$ at the LHC with $\sqrt{s} =$ 14 TeV are shown in Table~\ref{Xsections}, calculated at next-to-leading order (NLO) by means of the code {\tt HPAIR}~\cite{Spira:1997dg,Spira:2016ztx,Spira:website}, for $M_A =$ 200, 300, and 400 GeV, with $\tan\beta =$ 3. The cross sections predicted for the $q \bar q$ mode are sizable, even for the {\it heavy-mass} scenario with $M_A =$ 400 GeV, of ${\cal O}(1)$ fb at least. In contrast, the values for the $gg$ production mode are much smaller, between one and two orders of magnitude lower than $q \bar q \to HA$\footnote{The impact of the $gg$ mode on the production cross section of the signal results in a percent-level increase within the {\it moderate-mass} scenario which we have chosen to design the search strategy.}. According to these numbers, it is clear that in the low-$\tan\beta$ regime, for the reference value of $\tan\beta =$ 3, it is a reliable approximation to consider only the mode $q \bar q \to HA$ in order to calculate the $H+A$ production cross section.

\renewcommand{\arraystretch}{1.4}
\begin{table}[t]
\begin{center}
\begin{tabular}{|c|c|}
\hline
$H (\to XX) + A(\to YY)$ & $\sigma(q \bar q \to HA) \times \text{BR}$ [fb] \\[1mm] \hline \hline 
$W^+ W^-(\to l\nu_l jj')+\chi\chi$ &  3.704 \\ \hline
$b \bar b + \chi\chi$ &  1.533 \\ \hline
$\tau^+\tau^- + \chi\chi$ &  0.128 \\ \hline
$ZZ (\to 4l) + \chi\chi$ &  0.005 \\ \hline
\end{tabular}
\caption{$\sigma(q \bar q \to HA) \times \text{BR}\,[\mathrm{fb}]$ of the main decay modes of the heavy neutral MSSM Higgs bosons in the low-$\tan\beta$ regime for the {\it light-mass} scenario with $\sqrt{s} =$ 14 TeV.}
\label{ratesLM}
\end{center}
\end{table}

\begin{table}[t]
\begin{center}
\begin{tabular}{|c|c|}
\hline
$H(\to XX) + A(\to YY)$ & $\sigma(q \bar q \to HA) \times \text{BR}$ [fb] \\[1mm] \hline \hline 
$hh (\to 4b) +\chi\chi$ &  0.453 \\ \hline
$b \bar b + \chi\chi$ &  0.275 \\ \hline
$W^+ W^-(\to l\nu_l jj') + \chi\chi$ &  0.160 \\ \hline
$hh (\to 2b2\tau) +\chi\chi$ &  0.049 \\ \hline
\end{tabular}
\caption{$\sigma(q \bar q \to HA) \times \text{BR}\,[\mathrm{fb}]$ of the main decay modes of the heavy neutral MSSM Higgs bosons in the low-$\tan\beta$ regime for the {\it moderate-mass} scenario with $\sqrt{s} =$ 14 TeV.}
\label{ratesMM}
\end{center}
\end{table}

\begin{table}[t]
\begin{center}
\begin{tabular}{|c|c|}
\hline
$H(\to XX) + A(\to YY)$ & $\sigma(q \bar q \to HA) \times \text{BR}$ [fb] \\[1mm] \hline \hline
$t \bar t + t \bar t$ &  0.988 \\ \hline
$hh (\to 4b) + t \bar t $ & 0.067 \\ \hline
$t \bar t + \chi\chi$ & 0.043 \\ \hline
\end{tabular}
\caption{$\sigma(q \bar q \to HA) \times \text{BR}\,[\mathrm{fb}]$ of the main decay modes of the heavy neutral MSSM Higgs bosons in the low-$\tan\beta$ regime for the {\it heavy-mass} scenario with $\sqrt{s} =$ 14 TeV.}
\label{ratesHM}
\end{center}
\end{table}

In what follows we shall discuss the cross sections of the main decay channels for the $H+A$ pair production through the mode $q \bar q \to HA$. In order to compute the event rate for each decay mode, we use the branching ratios of $H$ and $A$ listed in Table~\ref{BRs} along with the $H+A$ production cross sections obtained with {\tt HPAIR}~\cite{Spira:1997dg,Spira:2016ztx,Spira:website} (see Table~\ref{Xsections}). Thus, the cross sections are given by $\sigma(pp \to Z^\ast \to HA) \times \text{BR}(H \to XX, A \to YY)$. As we want that one of the two heavy Higgs bosons decays invisibly, we will consider only the invisible channel of the pseudoscalar, $A \to \chi\chi$, since $A$ has the largest invisible rates, being in addition its dominant decay mode in the two first benchmarks. For those decay chains that include $H\to W^+ W^-$ we consider only the semileptonic decay mode, $W^+W^- \to l\nu_l+ jj$, with $l=e,\mu$, so that $\text{BR}(WW \to l\nu_l+ jj)= 0.29$. In the case of the decay $H \to ZZ$, we take into account the leptonic $Z$-boson decays, and thus $\text{BR}(ZZ \to 4l) \simeq 4 \times 10^{-4}$, with $4l = 4e, 2e2\mu, 4\mu$. For the processes with the $H \to hh$ decay mode, we will consider BR($h \to b \bar b$) $\simeq$ 0.7 and BR($h \to \tau^+\tau^-$) $\simeq$ 0.07. The resulting cross sections with $\sqrt{s} =$ 14 TeV are shown in Tables~\ref{ratesLM},~\ref{ratesMM}, and~\ref{ratesHM} for the three scenarios {\it light-mass}, {\it moderate-mass}, and {\it heavy-mass}, respectively. Some comments are required in each case:
\begin{itemize}
\item In the case of $M_A=200$ GeV ({\it light-mass} scenario), we see from Table~\ref{ratesLM} that the dominant decay mode is the final state $W^+ W^-(\to l\nu_l jj')+\chi\chi$, being the subdominant decay mode $b \bar b + \chi\chi$. The former has the advantage of a very large cross section, but the presence of $E_T^\text{miss}$ in one of the $W$ decay would result in an involved analysis since it affects the reconstruction of the missing transverse energy profile of the invisible $A$ decay, possibly preventing an efficient discrimination between signal and background. Moreover, the hadronic $W$ decay would need to be confronted with large QCD backgrounds. On the other hand, the latter seems to be more interesting, but a detailed study of this $b \bar b + E_T^\text{miss}$ channel~\cite{Banerjee:2016nzb} has demonstrated that it will be challenging to probe it, even during the HL-LHC run. Instead of the $H$ decay into a bottom-quark pair, the decay into a $\tau$-lepton pair is more efficient in reducing the QCD background, but it is affected by its relatively small cross section which is one order of magnitude lower than in $b \bar b$ channel. Moreover, the $\tau$-tagging efficiency is not much better than the $b$-tagging one. Finally, the cleanest channel would be $4l + E_T^\text{miss}$, where the four leptons come from the heavy scalar decay $H \to ZZ$ with both $Z$ bosons decaying leptonically. Unfortunately, this channel has a very tiny cross section and it would be a difficult challenge to enhance the signal significance over the SM backgrounds.
\item The final states allowed for the {\it moderate-mass} scenario seem to be more promising, as we can see in Table~\ref{ratesMM}, since the $H \to hh$\footnote{The resonant di-Higgs production has been widely studied in the MSSM (see for example~\cite{Dolan:2012ac}) and in many other models with an extended Higgs sector, as for example in contexts of strongly interacting Higgs sector~\cite{Abe:2012fb}, Higgs portal models~\cite{Dolan:2012ac,No:2013wsa}, and 2HDM~\cite{Lu:2015qqa,Ren:2017jbg}. However, this di-Higgs production $H \to hh$ has not been considered in any of them in association with a pseudoscalar Higgs boson decaying invisibly.} decay is open. Indeed, the dominant channel is $4b + E_T^\text{miss}$, in which the four bottom quarks are the decay products of two SM-like Higgs bosons $h$ originated from this decay. This class of final state, with another different decay chain, is also commented in~\cite{Banerjee:2016nzb} as a very promising channel to probe new physics at the LHC, but was not analyzed in detail. We could also take advantage of the presence of the decay $H \to hh$ and consider the channel $2b2\tau + E_T^\text{miss}$. Nevertheless, the cross section in this case is one order of magnitude lower than the $4b + E_T^\text{miss}$ final state, and we would have to deal again with large QCD backgrounds and $b$ and $\tau$ taggings anyway. In this kinematic region, the channels $W^+ W^- + E_T^\text{miss}$ and $b \bar b + E_T^\text{miss}$ still have important cross sections, but lower than the $4b + E_T^\text{miss}$ final state. However, these decay channels have the limitations already discussed in the {\it light-mass} scenario.
\item The kinematic region in which the threshold of $t \bar t$ production in the heavy Higgs decays is open ({\it heavy-mass} scenario) is dominated by far by the $4t$ final state (see Table~\ref{ratesHM}), with a cross section of $\sim$ 1 fb. The largest channel with missing transverse energy is $t \bar t + E_T^\text{miss}$, with a cross section 20 times smaller than the $4t$ channel. With such a low cross section the QCD background would overwhelm the signal in the case of hadronic top decays; whereas for leptonic top decays, the $E_T^\text{miss}$ coming from the tops would distort the missing transverse energy distribution from the invisible $A$ decay. Therefore, it seems rather complicated to probe the invisible Higgs decays through the heavy Higgs-pair production in this kinematic region. However, one can just study the heavy Higgs-pair production through the four tops channel, which could be detectable at the LHC with the search strategies designed in~\cite{Alvarez:2016nrz}. For this scenario, the branching ratio of $H \to hh$ keeps still a sizable value (11\%). Thus, another channel to try to probe the heavy Higgs-pair production could be $4b+2t$. Regrettably, the cross section is more than one order of magnitude lower than the $4t$ channel and the QCD background is also very large to obtain a substantial significance.
\end{itemize}
Taking into account the arguments stated above, we shall focus on the {\it moderate-mass} scenario, with the channel $q \bar q \to HA \to 4b+E_T^{\text{miss}}$ ($H \to hh \to 4b$ and $A \to \chi\chi$) as the most promising signal to probe invisible Higgs decays through heavy Higgs-pair production at the LHC. This same final state has been recently analyzed by CMS~\cite{Sirunyan:2017obz} within the context of gauge-mediated SUSY breaking, considering the electroweak production of two higgsinos that decay into the lightest Higgs boson $h$ and the LSP goldstino $\tilde G$. The results reported by CMS are consistent with the SM background predictions and 95\% CL exclusion limits are imposed on the higgsino-pair production cross sections, which are much larger than ours. Moreover, this $4b + E_T^\text{miss}$ final state, originated from this SUSY cascade process, has a very different kinematic behavior from the final state coming from our proposed signal $pp \to HA \to 4b+E_T^{\text{miss}}$. In the next section we will develop a dedicated search strategy for this channel. Before that, we show below that the recent LHC searches do not exclude a signal compatible with the production process $pp\to Z^*\to HA$ in any of the proposed benchmarks.

\subsection{LHC searches}

In order to probe the three scenarios introduced in Table~\ref{BRs}, we have confronted them with the general searches of new physics at the LHC at 8 TeV and 13 TeV by using the software {\tt CheckMATE 2}~\cite{Dercks:2016npn}. We have simulated the process $pp\to Z^*\to HA$ using {\tt MadGraph 5}~\cite{Alwall:2014hca} and decayed the heavy neutral Higgs bosons through the branching ratios computed with the {\tt SUSY-HIT} package~\cite{Djouadi:2006bz}. The cross section of the $HA$ production was computed at NLO by using {\tt HPAIR}~\cite{Spira:1997dg,Spira:2016ztx,Spira:website}, which includes QCD and SUSY-QCD corrections taken from~\cite{Dawson:1998py} and~\cite{Agostini:2016vze}, respectively. Finally, we have implemented {\tt PYTHIA}~\cite{Sjostrand:2014zea} for the parton shower and hadronization, and the detector simulation was carried out using {\tt Delphes 3}~\cite{deFavereau:2013fsa}.

\begin{table}
\begin{center}
\begin{tabular}{|c|c|c|}
\hline
 $M_A [\mathrm{GeV}]$ & $S/\sqrt{B}$ LHC at 8 TeV (13 TeV) & Signal Region \\[1mm] \hline \hline 
$200$  & $0.058$ $(0.077)$ & S7~\cite{Aad:2014qaa} (bCbv~\cite{ATLAS:2016ljb})  \\ \hline
$300$  & $0.029$ $(0.022)$ & SR-0l-4j-A~\cite{TheATLAScollaboration:2013tha} (bCbv~\cite{ATLAS:2016ljb})  \\ \hline
$400$  & $0.025$ $(0.016)$ & SR3b~\cite{Aad:2014pda} (SR3b~\cite{Aad:2016tuk})  \\ \hline
\end{tabular}
\label{sig1}
\caption{Significances corresponding to the LHC at 8 TeV and 13 TeV (in parentheses) obtained from {\tt CheckMATE 2}, for the three benchmarks with $M_A=$ 200, 300, and 400 GeV. In the last column the name of the most restrictive signal region is included.}
\label{sigCheckMATE}
\end{center}
\end{table}

For both the 8 TeV and the 13 TeV searches, none of the three scenarios is excluded by the {\tt CheckMATE 2} validated analyses. For each analysis implemented in {\tt CheckMATE 2}, we have also computed the signal significance of the most restrictive signal region. We have used the approximate formula $S/\sqrt{B}$, where $S$ is the number of signal events in a given signal region and $B$ is the corresponding number of background events. The results derived from the LHC searches at 8 TeV included in {\tt CheckMATE 2} are listed in Table~\ref{sigCheckMATE}. In the last column, we include the name of the most restrictive signal region as given in the corresponding experimental analyses. For the three scenarios, the values of $S/\sqrt{B}$ obtained with the signal regions of the general new physics LHC searches at 8 TeV are low. The results corresponding to the LHC searches at 13 TeV are also shown in Table~\ref{sigCheckMATE} in parentheses. As in the case of the 8 TeV searches, the values obtained for the significance are also low for the three scenarios.

For $M_A=200$ GeV, the most restrictive signal regions are S7 and bCbv for the LHC searches at 8 and 13 TeV, respectively. The former is defined in the analysis associated to the ATLAS search for direct top-squark pair production in final states with two leptons of opposite charge using $20.3~\mbox{fb}^{-1}$~\cite{Aad:2014qaa}, while the latter corresponds to the ATLAS search for top squarks in final states with one isolated lepton, jets, and missing transverse momentum using $13.2~\mbox{fb}^{-1}$~\cite{ATLAS:2016ljb}. The signal region S7 is defined through the lepton-based stransverse mass, $m_{T2}$, requiring $m_{T2}>120$ GeV and also in such a way that the number of jets with $p_T>20$ GeV must be smaller than 2. On the other hand, the signal region bCbv selects events with two or more jets with $p_T> (120~\mbox{GeV},80~\mbox{GeV})$ and no $b$-tagged jets. The following requirements are also imposed: $E^{\mathrm{miss}}_T>360$ GeV, $H^{\mathrm{miss}}_{T,sig}>16$, $m_T>200$ GeV, $|\Delta\phi(\text{jet}_i,\vec{p}^{\,\,\mathrm{miss}}_T)|(i=1)>2.0$, $|\Delta\phi(\text{jet}_i,\vec{p}^{\,\,\mathrm{miss}}_T)|(i=2)>0.8$, leading large-R jet mass [70 GeV, 100 GeV], and $\Delta\phi(\vec{p}^{\,\,\mathrm{miss}}_T,\ell)>1.2$. The definition of the different variables can be found in~\cite{ATLAS:2016ljb}.

In the case of $M_A=300$ GeV, the signal regions with the highest significance are SR-0l-4j-A~\cite{TheATLAScollaboration:2013tha} and again bCbv~\cite{ATLAS:2016ljb} for 8 and 13 TeV, respectively. The signal region SR-0l-4j-A is included in the analysis of the ATLAS search for strong production of SUSY particles in final states with missing transverse momentum and at least three $b$-jets using $20.1~\mathrm{fb}^{-1}$~\cite{TheATLAScollaboration:2013tha}. The specific selection for this signal region is based on the requirements: $N_{\mathrm{jets}}>4$, $p_T$ of jets $> 30$ GeV, $E^{\mathrm{miss}}_T>200$ GeV, $m^{4j}_{\mathrm{eff}}>1000$ GeV, and $E^{\mathrm{miss}}_T/\sqrt{H^{4j}_T}>16~\mathrm{GeV}^{1/2}$. The definition of these variables are given in~\cite{TheATLAScollaboration:2013tha}.

Finally, for $M_A=400$ GeV the signal region that exhibits the highest significance is called SR3b for both cases of 8 TeV and 13 TeV searches, although its definition is different for each analysis. In both cases, the search is focused on final states with jets and two same sign (SS) leptons or three leptons. For the 8 TeV analysis~\cite{Aad:2014pda}, this signal region requires two SS leptons or three leptons with at least five jets and at least three $b$-jets, and also $m_{\mathrm{eff}}>350$ GeV, while for the 13 TeV analysis~\cite{Aad:2016tuk} the requirements becomes: $N_{\mathrm{lept}}\geq 2$, at least three $b$-jets with $p_T>20$ GeV, $E_T^\text{miss}>125$ GeV, and $m_{\mathrm{eff}}>650$ GeV.

The analyses considered above show that even the most restrictive signal regions corresponding to general searches are far from being sensitive to the heavy Higgs-pair production. It is then sensible to design a dedicated search strategy for this process. We present in the next section a detailed discussion on this matter.

\section{Search strategy for \texorpdfstring{$pp \to HA \to 4b+E_T^{\text{miss}}$}{qqHA} at the LHC}
\label{strategy}
In this section we develop a search strategy for the production of a pair of heavy Higgs bosons, $H$ and $A$, with $M_H =$ 309 GeV and $M_A =$ 300 GeV ({\em moderate mass} scenario), and the subsequent decays of the pseudoscalar Higgs boson into its invisible channel and the heavy scalar into a pair of light Higgs bosons $h$, which decay into a bottom-quark pair. The final state resulting from the chosen signal process contains 4 $b$-jets and large $E_T^{\text{miss}}$, with a total cross section of 0.453 fb. For this final state topology, we have considered a first signal region, SR1, defined by requiring exactly 4 $b$-tagged jets. However, since in our working point the maximum $b$-tagging efficiency is $\sim 75\%$, we have also studied a second signal region, SR2, in which the above $b$-tagging requirement is relaxed to 3 or 4 $b$-tagged jets. In addition, both signal regions require 0 or 1 light-jet to suppress multi-jet backgrounds, and include a lepton veto to disfavor the presence of missing transverse energy due to neutrinos in backgrounds involving top quarks. In what follows, we present first the general features of the signal and a procedure to optimize its potential detection. Later on, we will detail the search strategies for both signal regions.

\subsection{General signal features and cut optimization procedure}
\label{signal-charact}

Let us start remembering that the proposed signal final state is made up of 4 $b$-jets and it has large $E_T^\text{miss}$, having a total cross section of 0.453 fb, and stating that the coming discussion on the procedure used to maximize the sensitivity of the signal will be centered on the irreducible backgrounds, postponing our treatment of the reducible contributions to Sections~\ref{4b_search} and~\ref{3y4b_search}. The motivation is that the former expectedly follow the signal more closely than the latter. Thus, an optimization of the signal significance based on the irreducible backgrounds may be sufficient to achieve an efficient discrimination among the signal and all background sources. In other words, we first develop the search strategy by taking into account the signal and only the irreducible backgrounds, and subsequently, we estimate the reducible contribution. The main irreducible SM backgrounds for this final state, in both signal regions SR1 and SR2, are $t \bar t b \bar b$, $Z(\to \nu\bar\nu) b \bar b b \bar b$, and $Z(\to \nu\bar\nu) b \bar b b \bar b j$. Both the signal and the main irreducible backgrounds have been generated with {\tt MadGraph\_aMC@NLO}~\cite{Alwall:2014hca} and showered with {\tt PYTHIA}~\cite{Sjostrand:2014zea}, simulating the detector response with {\tt Delphes 3}~\cite{deFavereau:2013fsa}. The events have been generated within the four-flavor scheme with the factorization and renormalization scales set to $\mu_R = \mu_F = H_T/2$ and using the PDF MSTW2008nlo68cl\_nf4. Besides, the following {\it basic cuts} have been imposed at generator level\footnote{These cuts are consistent with the usual selection performed by the trigger system and also by the reconstruction algorithms in the experimental analyses. Since we have checked that the distortions on these kinematic variables due to the parton shower are statistically negligible, we apply these cuts at the generator level, which makes the simulation process more efficient. Finally, note that the same cuts are applied both to light and b-jets, which prevents the impact of potential misidentification at detector level on the corresponding kinematic distributions.}
\begin{eqnarray}
&& p_T^b > 30 \, \text{GeV} \,\,,\,\,  |\eta_b| < 2 \,, \nonumber\\  
&& p_T^{j} > 30 \, \text{GeV}  \,\,,\,\, |\eta_{j}| < 2 \,,  \nonumber\\ 
&& \Delta R _{bb} > 0.2 \,\,,\,\, \Delta R _{bj} > 0.2 \, .\qquad \nonumber
\end{eqnarray}
For the signal and each one of the main irreducible backgrounds, the number of generated events has been much larger than the expected amounts for an integrated luminosity of $\sim 1500\,$fb$^{-1}$ and a center-of-mass energy of $\sqrt{s}= 14\,$TeV\footnote{Due to the rather large cross section, the number of generated events for the $t \bar t b \bar b$ background has been five times larger than the amount of events expected at 300 fb$^{-1}$.}. Let us remember that the signal cross section has been obtained at NLO by means of the codes {\tt HPAIR} and {\tt SUSY-HIT} (see Section~\ref{HAproduction}), whilst for the backgrounds we have used the NLO cross section calculated with {\tt MadGraph\_aMC@NLO} only in the case of $t\bar t b \bar b$. The cross sections obtained with {\tt MadGraph\_aMC@NLO} for the three main irreducible backgrounds are given by $$\sigma(t\bar t b \bar b)=1633~\text{fb},\,\sigma(Z(\to \nu\bar\nu) b \bar b b \bar b)=3.27~\text{fb},\,\sigma(Z(\to \nu\bar\nu) b \bar b b \bar b j)=2.45~\text{fb}\,.$$ 

After the event generation with the {\it basic cuts}, the next step on the analysis will be to require that both the signal and background events pass a set of {\it selection cuts} at detector level, whose definition depends on the signal region:
\begin{itemize}
\item SR1: 4 $b$-tagged jets, 0 leptons, 0 or 1 light jet, along with the requirement $p_{T}^{\text{leading-}b}>$ 70 GeV on the transverse momentum of the leading $b$-jet.
\item SR2: 3 or 4 $b$-tagged jets, 0 leptons, 0 or 1 light jet, and require also the transverse momentum of the third leading $b$-jet to be above 30 GeV.
\end{itemize}

\begin{figure}[t!]
\begin{center}
\begin{tabular}{cc}
\includegraphics[width=80mm]{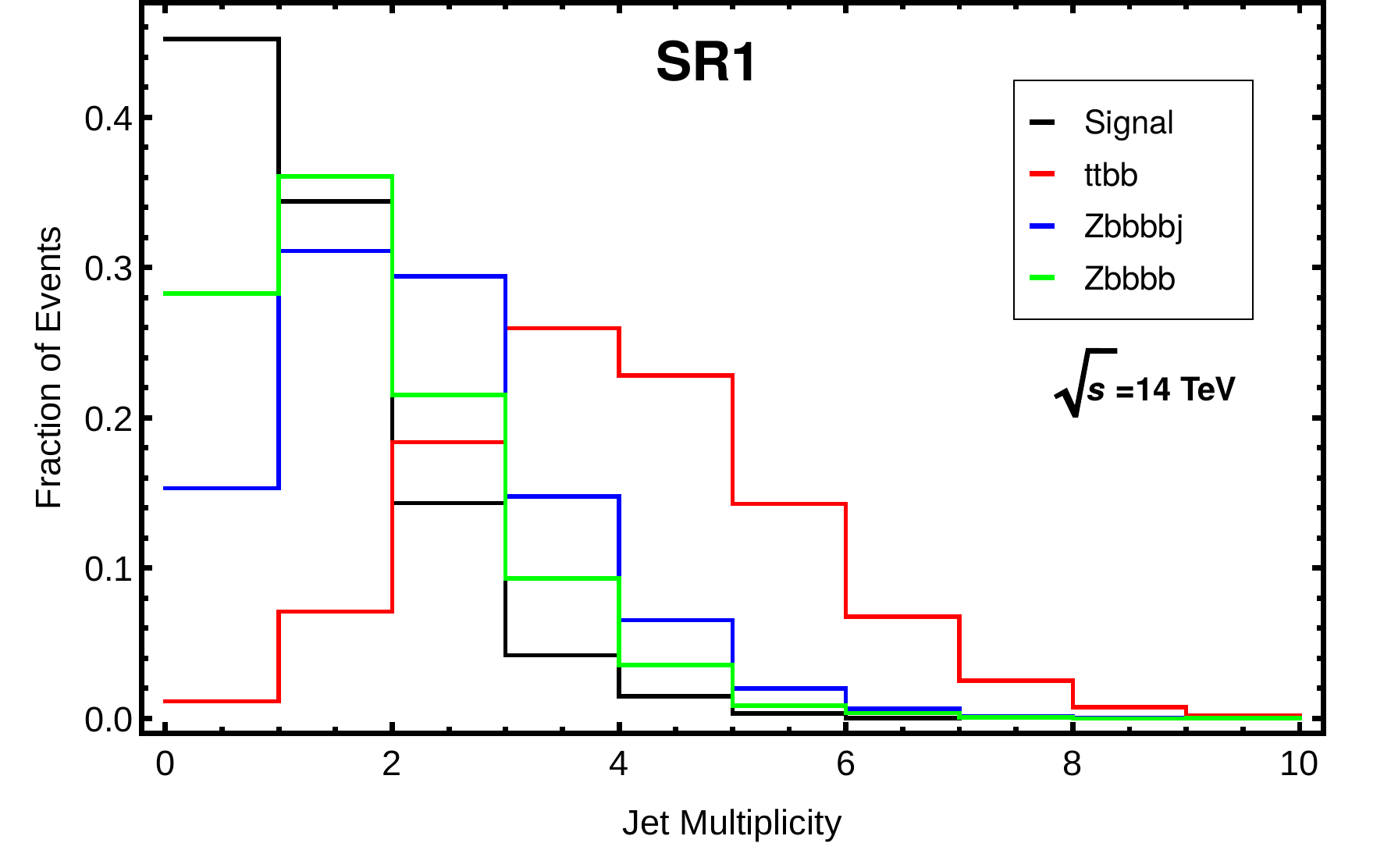} &
\includegraphics[width=80mm]{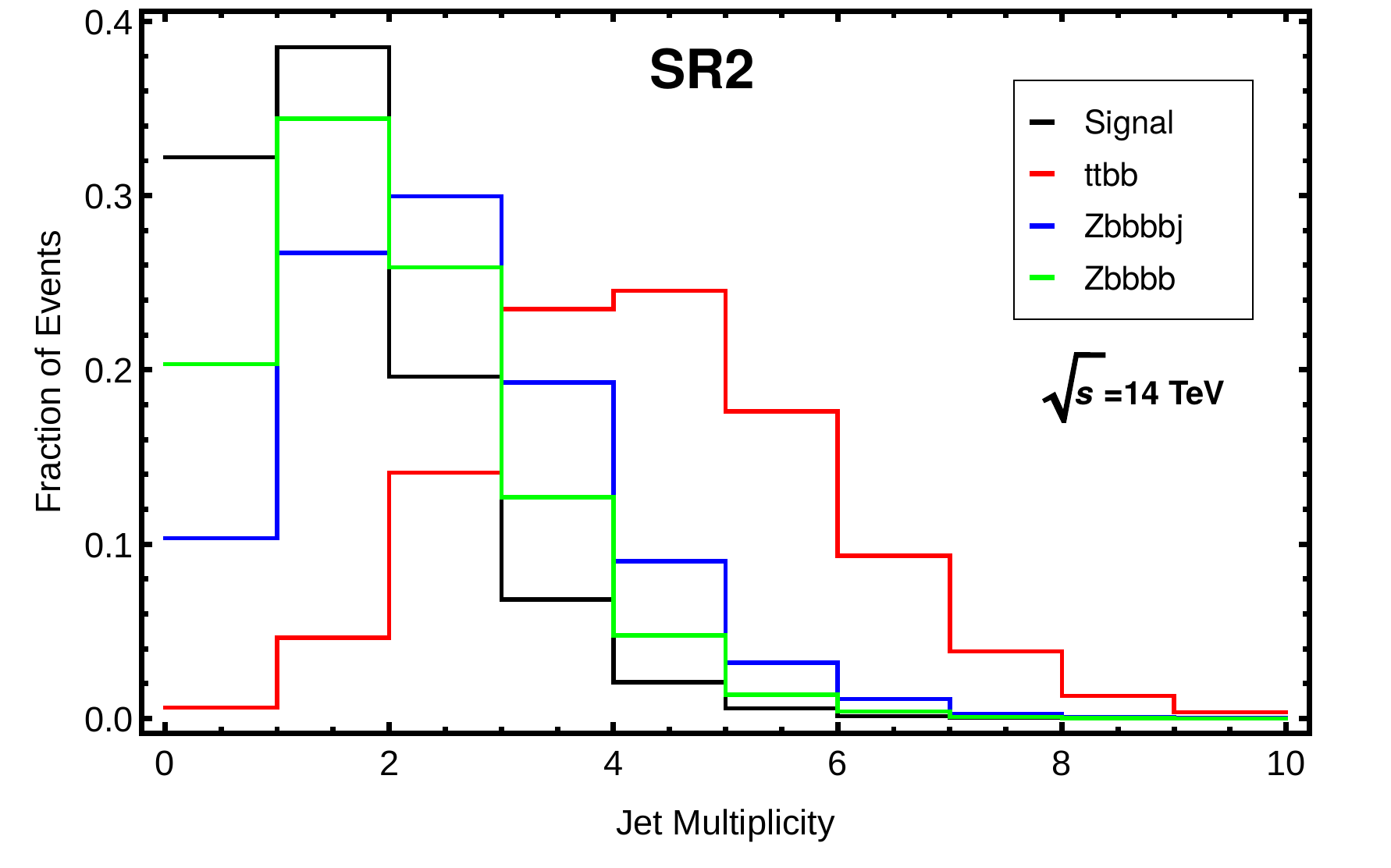}
\end{tabular}
\caption{Distribution of the number of light jets after the {\it selection cuts} without the light-jet requirement for the signal and for the main irreducible backgrounds with $\sqrt{s}$ = 14 TeV in the signal regions SR1 (left panel) and SR2 (right panel).}\label{Nlightjets_distributions}
\end{center} 
\end{figure}

The jet multiplicity of the backgrounds is expected to be much larger than the signal one, as it can be observed in Figure~\ref{Nlightjets_distributions}, in which we display the distribution of the number of light jets after the {\it selection cuts}, without the light-jet requirement, for the signal and the main irreducible backgrounds in the signal regions SR1 (left panel) and SR2 (right panel). From these two plots it is clear that a cut in the number of light jets, demanding 0 or 1 light jet at most, will keep a large part of the signal while removing a considerable portion of the backgrounds rates, especially for $t \bar t b \bar b$, which has the largest cross section by far.

\begin{figure}[t!]
\begin{center}
\begin{tabular}{cc}
\includegraphics[width=80mm]{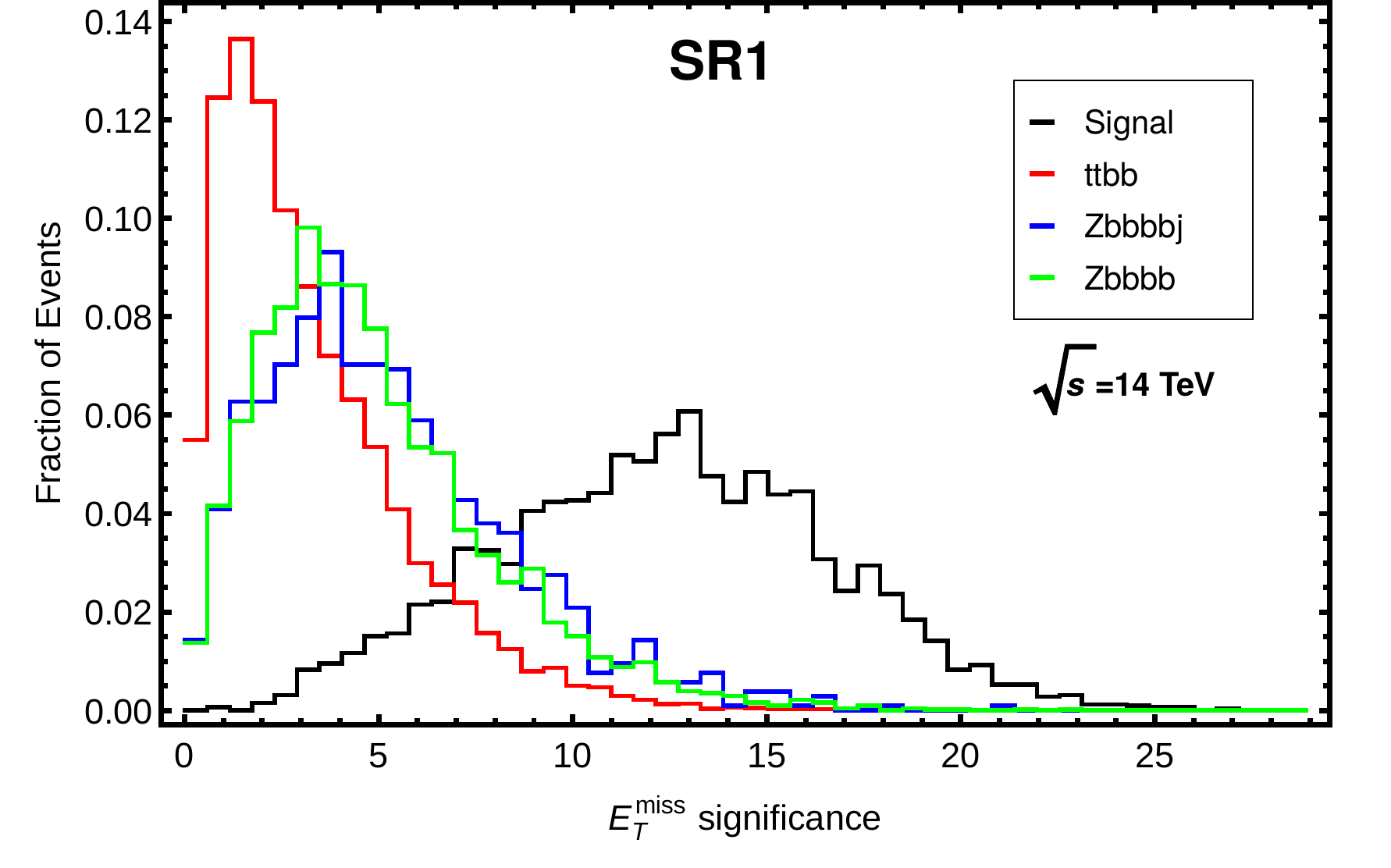} &
\includegraphics[width=80mm]{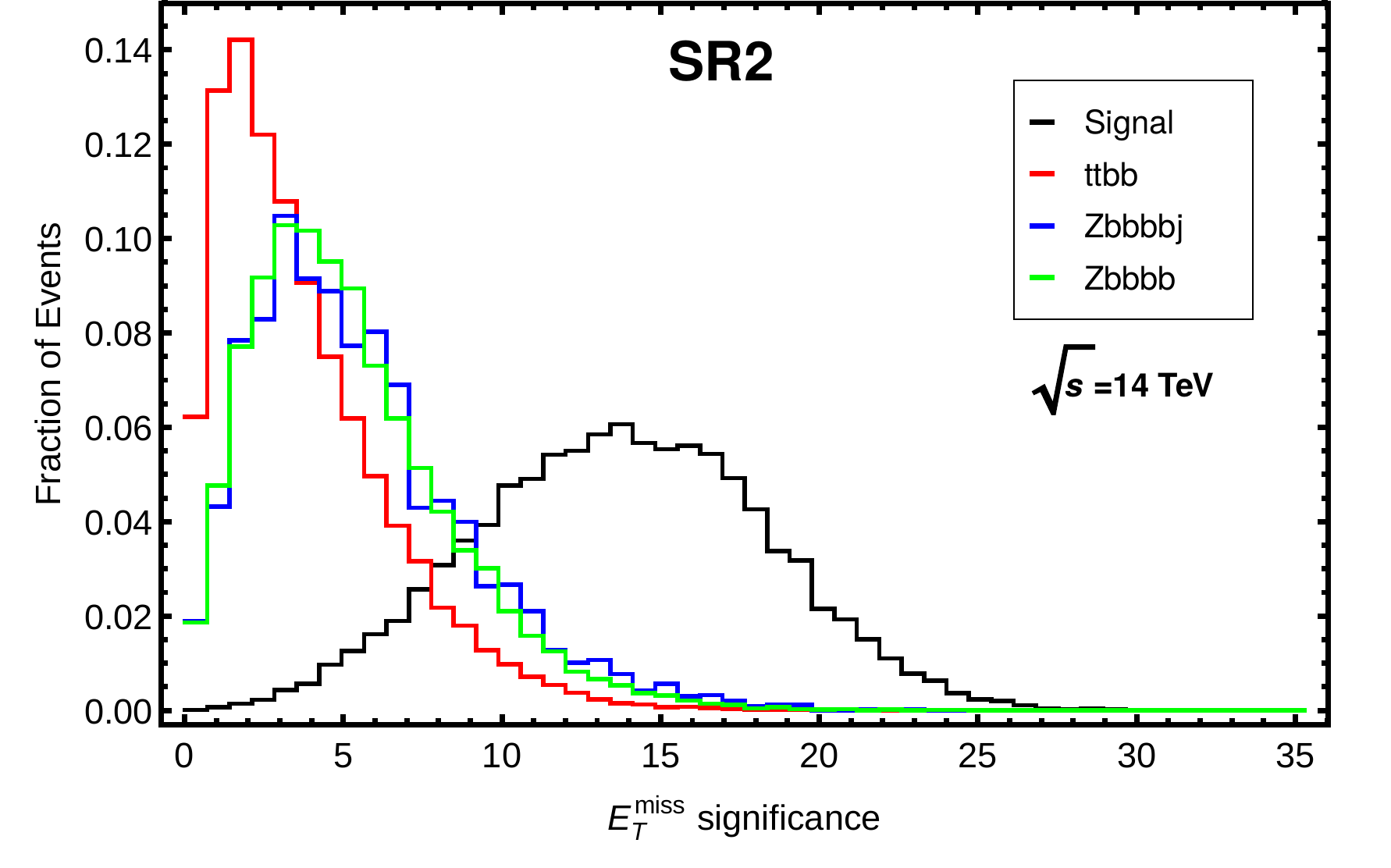}
\end{tabular}
\caption{Distribution of the $E_T^\text{miss}$ significance after the {\it selection cuts} for the signal and for the main irreducible backgrounds with $\sqrt{s}$ = 14 TeV in the signal regions SR1 (left panel) and SR2 (right panel).}\label{Etmiss-sig_distributions}
\end{center} 
\end{figure}

Another interesting feature of the signal is the expected very large values of $E_T^\text{miss}$ as compared with the backgrounds. Therefore, a potential powerful discriminator between the signal and backgrounds is the variable $E_T^\text{miss}$ significance, defined as the ratio $E_T^{\text{miss}} / \sqrt{H_{T}^{nj}}$, where $H_{T}^{nj}=\sum_i p^{i}_{T}$, with the sum running over the $n$ leading jets considered in each signal region. More specifically, we have set $n=$ 4 and 3 for SR1 and SR2, respectively. The distribution of this $E_T^\text{miss}$ significance is depicted in Figure~\ref{Etmiss-sig_distributions} in both signal regions, for the signal and main backgrounds, after the {\it selection cuts}. The corresponding distributions for the signal and the backgrounds are very different, with central values of $E_T^{\text{miss}} / \sqrt{H_{T}^{nj}} <$ 5 GeV$^{1/2}$ for the latter and $E_T^{\text{miss}} / \sqrt{H_{T}^{nj}} >$ 10 GeV$^{1/2}$ for the former, which points out that a cut requiring values of the $E_T^\text{miss}$ significance larger than 10 GeV$^{1/2}$ would largely increase the signal-over-background ratio. Besides, in contrast to the $E_T^\text{miss}$ alone, the $E_T^\text{miss}$ significance leads to a better discrimination even  among the different backgrounds allowing in turn a higher significance of the signal. In particular, a cut on $E_T^\text{miss}$ significance will have a considerable impact on the $t \bar t b \bar b$ background which after imposing the {\it selection cuts} is characterized by events with instrumental $E_T^\text{miss}$ and, precisely, the $E_T^\text{miss}$ significance is highly efficient to discriminate fake from genuine $E_T^\text{miss}$.

Apart from the cuts described above, which are common for both signal regions SR1 and SR2, we have also considered a wide set of angular variables (more than 40 altogether), related to the azimuthal angle $\phi$ and the pseudorapidity $\eta$. These angular variables have been introduced with the aim to amplify the differences between the signal and background distributions, which still have similar kinematics after the imposition of: the {\it selection cuts}, the restrictions over the $E_T^\text{miss}$ significance, and over the invariant mass variables (only in the case of SR1). As we will show in the next section, the use of some of the angular variables turns out to be useful to improve the significance of the signal. 

The procedure to select the optimal values of the cuts for the different kinematic variables has been developed by means of a sequential optimization of the statistical significance, $\mathcal{S}$, calculated according to the expression~\cite{Cowan:2010js}:
\begin{equation}
\label{Sig1}
\mathcal{S}=\sqrt{-2 \left((B+S) \log \left(\frac{B}{B+S}\right)+S\right)} \,,
\end{equation}
where $S$ and $B$ denote the number of signal and background events, respectively. The ultimate goal of the cut optimization is to determine a search strategy that maximizes the statistical significance in each signal region. The procedure starts with the selection of events that pass the discrete cuts on the number of $b$-jets, light-jets, and leptons. After that, the distributions of the kinematic variables (transverse momenta of the final state particles, $E_T^\text{miss}$ significance, invariant masses, angular variables) are then scrutinized. Once the variable that shows a better discriminating power between signal and background is chosen, the optimization algorithm is executed just for this variable and the optimal value of the cut is applied to the events. This procedure is then repeated by picking and optimizing the most discriminating variable at each step until the maximum  statistical significance is achieved, keeping at least one signal event. Finally, the acceptances for the signal and backgrounds, defined as the proportion of events at the detector level that pass the cuts implemented at each step with respect to the amount of events generated with the {\it basic cuts}, are used to apply correction factors to the expected number of events at a certain luminosity.

\subsection{SR1: 4 $b$-jets signal region}
\label{4b_search}
In this case, besides the main irreducible backgrounds $t \bar t b \bar b$, $Z(\to \nu\bar\nu) b \bar b b \bar b$, and $Z(\to \nu\bar\nu) b \bar b b \bar b j$ mentioned above, we have to deal with the major reducible backgrounds that arise from $t\bar t$, $t\bar t\, +\,$jets and $Z\, +\,$jets. Specifically, we have considered
$$t \bar t,\, t \bar t j,\, t \bar t j j\, ,$$
\vspace*{-6mm}
$$Z b \bar b j j,\, Z j j j j\, ,$$
\vspace*{-6mm}
\begin{equation}
\nonumber
Z b \bar b j j j,\, Z j j j j j\,\footnote{There is an additional source of reducible background that we have not included in our analysis arising from multi-jet production. The cross section for this background is huge and a data driven approach is necessary. However, since in this case there is not a genuine source of missing transverse energy but just a fake contribution as a consequence of imperfect reconstruction of objets and energy resolution, an $E_T^\text{miss}$ significance $\sim$ 1 GeV$^{1/2}$ is expected for this background, much below the value of the cut applied to this variable, as we will see below.}.
\end{equation}
With the setup described in Section~\ref{signal-charact}, the cross sections for the reducible backgrounds obtained with {\tt MadGraph\_aMC@NLO} are given by
$$\sigma(t \bar t)=5.17\times 10^5~\text{fb}\,,\,\sigma(t \bar t j)=2.25\times 10^5~\text{fb}\,,\, \sigma(t \bar t j j)=8.72\times 10^4~\text{fb},$$
\vspace*{-6mm}
$$ \sigma(Z(\to \nu\bar\nu) b \bar b j j)= 9.60\times 10^2~\text{fb}\,,\, \sigma(Z(\to \nu\bar\nu) j j j j)=1.48\times 10^4~\text{fb},$$
\vspace*{-5mm}
\begin{equation}
\nonumber
 \sigma(Z(\to \nu\bar\nu) b \bar b j j j)=3.22\times 10^2~\text{fb}\,,\, \sigma(Z(\to \nu\bar\nu) j j j j j)=1.42\times 10^4~\text{fb}
 \vspace*{2mm} \,.
 \end{equation}
 \par
The search strategy developed for this signal region proceeds trough the following steps:
\begin{itemize}
\item[(a)] We apply the {\it selection cuts} introduced in Section~\ref{signal-charact}, namely, 4 $b$-tagged jets, 0 leptons, 0 or 1 light-jet, along with the requirement $p_{T}^{\text{leading-}b}>$ 70 GeV on the transverse momentum of the leading $b$-jet.
\item[(b)] We use the $E_T^{\text{miss}}$ significance defined as the ratio $E_T^{\text{miss}} / \sqrt{H_{T}^{4j}}$, where $H_{T}^{4j}=\sum_i p^{i}_{T}$, with the sum running over the four leading jets. In particular, we set the cut $E_T^{\text{miss}} / \sqrt{H_{T}^{4j}}>13 \mbox{ GeV}^{1/2}$.
\item[(c)] We require the four $b$-tagged jets to be consistent with the decay chain $H \to hh \to b \bar b b \bar b$. More precisely, we restrict the search to the region 250 GeV $<m_{4b}<$ 340 GeV, where $m_{4b}$ is the invariant mass of the four $b$-jets, and also make use of the variable $\chi_{hh}$~\cite{Aad:2015uka} defined as:
 $$\chi_{hh}= \sqrt{\left(\frac{M_h-m_{b_1b_2}}{0.1 M_h}\right)^{2}+\left(\frac{M_h-m_{b_3b_4}}{0.1 M_h}\right)^{2}} \,,$$
 requiring that over all possible combinations of $b_1$-$b_4$ at least one of them gives a $\chi_{hh}$ value below 2.7. This last cut on $\chi_{hh}$ is intended to select events in which at least two disjoint pairs of $b$-jets are most likely to arise from the decay $h \to b \bar b$.
 \item[(d)] We perform a cut on the azimuthal angular separation between a $b$-jet and the missing transverse energy. Among all the possible values, we require the second minimum between any $b$-jet and $E_T^\text{miss}$ in the event ($\Delta \phi_{bE_{T}^{\text{miss}}}^{2^\text{nd}\text{min}}$) to be above 1.6.
\end{itemize}

\begin{table}
\begin{center}
\begin{tabular}{rrrrrr|c}
\hline \hline
Process & Signal & $t \bar t b \bar b$ & $Z b \bar b b \bar b$ & $Z b \bar b b \bar b j$ & Reducibles & ${\cal S}$  \\ \hline
Expected & 137 & 489900 & 981 & 734 & 9.1$\times 10^{6}$ & 0.04 \\ \hline
Selection cuts & 4.31 & 2532.20 & 42.44 & 23.39 & 45.0 & 0.08 \\
$E^{\text{miss}}_T / \sqrt{H_{T}^{4j}}$ & 1.95 & 8.62 & 0.82 & 0.7 & 1.09 & 0.56 \\
$m_{4b}$ + $\chi_{hh}$ & 1.08 & 0.2 & 0.03 & 0.07 & 0.08 & 1.33 \\
$\Delta \phi_{bE^{\text{miss}}_{T}}^{2^\text{nd}\text{min}}$ & 1.07 & 0 & 0.03 & 0.07 & 0.08 & 1.62 \\[1mm] \hline \hline
\end{tabular}
\caption{Signal and background event cut flow corresponding to the first signal region (SR1) for $\sqrt{s}$ = 14 TeV and a total integrated luminosity of ${\cal L}$ = 300 fb$^{-1}$.}
\label{cut-flow_300_SR1}
\end{center}
\end{table}

\begin{table}
\begin{center}
\begin{tabular}{rrrrrr|c}
\hline \hline
Process & Signal & $t \bar t b \bar b$ & $Z b \bar b b \bar b$ & $Z b \bar b b \bar b j$ & Reducibles & ${\cal S}$  \\ \hline
Expected & 686 & 2.4$\times 10^6$ & 4903 & 3671 & 4.5$\times 10^{7}$ & 0.1 \\ \hline
Selection cuts & 21.56 & 12661.0 & 212.2 & 117.0 & 225.0 & 0.19 \\
$E^{\text{miss}}_T / \sqrt{H_{T}^{4j}}$ & 9.74 & 43.11 & 4.12 & 3.50 & 5.45 & 1.26 \\
$m_{4b}$ + $\chi_{hh}$ & 5.41 & 0.98 & 0.15 & 0.38 & 0.42 & 2.97 \\
$\Delta \phi_{bE^{\text{miss}}_{T}}^{2^\text{nd}\text{min}}$ & 5.34 & 0 & 0.15 & 0.38 & 0.42 & 3.63 \\[1mm] \hline \hline
\end{tabular}
\caption{Signal and background event cut flow corresponding to the first signal region (SR1) for $\sqrt{s}$ = 14 TeV and a total integrated luminosity of ${\cal L}$ = 1500 fb$^{-1}$.}
\label{cut-flow_1500_SR1}
\end{center}
\end{table}

The cut flow for the signal and background events obtained by applying the steps described above is shown in Table~\ref{cut-flow_300_SR1} for a total integrated luminosity of ${\cal L}$ = 300 fb$^{-1}$. The cut flow is the result of the sequential optimization of the statistical significance, $\mathcal{S}$, as introduced in Section~\ref{signal-charact}. At each step we display the events corresponding to the signal and to the irreducible backgrounds along with the corresponding significance. Besides, we display in Table~\ref{cut-flow_1500_SR1} the cut flow corresponding to a total integrated luminosity of 1500 fb$^{-1}$. In this case no new optimization procedure has been applied because the acceptances can be safely rescaled, since the amount of generated events is consistent with the required luminosity. In addition, since the $Z\, + \,$jets reducible backgrounds are not attainable through our simulations, we show in the sixth column an estimation of the number of events corresponding to these reducible backgrounds that remain after imposing the cuts involved in each step. For this estimation, we have used conservative misidentification rates for light jets, $\epsilon_j=5\times 10^{-3}$, and $c$-jets, $\epsilon_c=0.15$, and a nominal $b$-tagging efficiency of $75\%$. For the steps (b) to (d), we have also assumed $Z\, + \,$jets to have the acceptances of $Z b \bar b b \bar b$ or $Z b \bar b b \bar b j$ (according to the number of jets in the final state), since similar kinematic distributions of events are potentially expected for this type of backgrounds. Finally, the $t \bar t \, + \,$jets reducible backgrounds are also beyond our computational resources to simulate them. However, the acceptances of the $t \bar t \, + \,$jets are smaller at each step of the cut flow than the acceptances of the corresponding irreducible background $t \bar t b \bar b$\footnote{In order to make this statement, we have generated $1 \times 10^6$ events of $t \bar t$ of which none of them passes the second step of the cut flow (step (b)). Thus we consider that an upper bound on the acceptance at this level is ${\cal O}(10^{-7})$, which is much lower than the acceptances for the $t \bar t b \bar b$ irreducible background.}. Therefore, it seems fairly sensible to assume that no $t \bar t \, + \,$jets events would survive at the end of this search strategy.

\subsection{SR2: 3 or 4 $b$-jets signal region}
\label{3y4b_search}

The main backgrounds in this signal region are the same as those involved in SR1, except for the addition of two new reducible backgrounds, $Z(\to \nu\bar\nu)bbj$ and $Z(\to \nu\bar\nu)jjj$, with cross sections of $2536~\text{fb}$ and $7564~\text{fb}$, respectively.  By using the same set of simulated events as in the case of SR1, the search strategy for SR2 involves the following steps:
\begin{itemize}
\item[(a)] We select events with 3 or 4 $b$-jets, 0 leptons, 0 or 1 light-jet, and also require the transverse momentum of the third leading $b$-jet to be above 30 GeV.
\item[(b)] We impose $E^{\text{miss}}_T / \sqrt{H_{T}^{3j}}>16.5\mbox{ GeV}^{1/2}$, where now $H_{T}^{3j}$ sums over the three leading jets.
\item[(c)] We apply two additional cuts: $\Delta \phi_{bb}^\text{max}<$ 1.90 and $\Delta \eta_{bb}^\text{max}<$ 1.42, where $\Delta \phi_{bb}^\text{max}$ is the maximum azimuthal angular separation between two $b$-jets and $\Delta \eta_{bb}^\text{max}$ is the maximum difference in pseudorapidities between the $b$-jets in the event.
\end{itemize} 

\begin{table}
\begin{center}
\begin{tabular}{rrrrrr|c}
\hline \hline
Process & Signal & $t \bar t b \bar b$ & $Z b \bar b b \bar b$ & $Z b \bar b b \bar b j$ & Reducibles & ${\cal S}$  \\ \hline
Expected & 137 & 489900 & 981 & 734 & 3.3$\times 10^{7}$ & 0.02 \\ \hline
Selection cuts & 21 & 6217 & 162.62 & 84.40 & 5825 & 0.19 \\
$E^{\text{miss}}_T / \sqrt{H_{T}^{3j}}$ & 5.83 & 7.05 & 0.68 & 0.70 & 27.23 & 0.95 \\
$\Delta \phi_{bb}^\text{max}$ + $\Delta \eta_{bb}^\text{max}$ & 4.19 & 0 & 0.14 & 0.12 & 5.38 & 1.59 \\ \hline \hline
\end{tabular}
\caption{Signal and background event cut flow corresponding to the second signal region (SR2) for $\sqrt{s}$ = 14 TeV and a total integrated luminosity of ${\cal L}$ = 300 fb$^{-1}$.}
\label{cut-flow_300_SR2}
\end{center}
\end{table}

\begin{table}
\begin{center}
\begin{tabular}{rrrrrr|c}
\hline \hline
Process & Signal & $t \bar t b \bar b$ & $Z b \bar b b \bar b$ & $Z b \bar b b \bar b j$ & Reducibles & ${\cal S}$  \\ \hline
Expected & 686 & 2.4$\times 10^{6}$ & 4903 & 3671 & 1.63$\times 10^{8}$ & 0.05 \\ \hline
Selection cuts & 105 & 31085 & 813 & 422 & 29126 & 0.42 \\
$E^{\text{miss}}_T / \sqrt{H_{T}^{3j}}$ & 29.17 & 35.27 & 3.38 & 3.50 & 136.14 & 2.13 \\
$\Delta \phi_{bb}^\text{max}$ + $\Delta \eta_{bb}^\text{max}$ & 20.93 & 0 & 0.69 & 0.62 & 26.89 & 3.56 \\ \hline \hline
\end{tabular}
\caption{Signal and background event cut flow corresponding to the second signal region (SR2) for $\sqrt{s}$ = 14 TeV and a total integrated luminosity of ${\cal L}$ = 1500 fb$^{-1}$.}
\label{cut-flow_1500_SR2}
\end{center}
\end{table}

The cut flow of the signal and background events obtained by applying the steps (a), (b), and (c) are shown in Tables~\ref{cut-flow_300_SR2} and \ref{cut-flow_1500_SR2}, again for total integrated luminosities of ${\cal L}$ = 300 fb$^{-1}$ and 1500 fb$^{-1}$, respectively. Following the same approach as in SR1, Table~\ref{cut-flow_300_SR2} has been obtained through the sequential optimization of the statistical significance whereas Table~\ref{cut-flow_1500_SR2} is a projection of the results at 300 fb$^{-1}$. We have dealt with the reducible backgrounds in a similar way as in SR1. They have been introduced into the cut flow through an estimation of their impact using the same misidentification rates for light jets ($\epsilon_j=5\times 10^{-3}$), $c$-jets ($\epsilon_c=0.15$), and the nominal $b$-tagging efficiency ($75\%$). For the steps (b) to (d), we have also assumed the $Z\, + \,$jets backgrounds to have the same acceptances corresponding to $Z b \bar b b \bar b$ or $Z b \bar b b \bar b j$. Finally, the $t \bar t \, + \,$jets backgrounds have received the same treatment as in SR1 since analogous arguments are valid in this signal region as well.

\subsection{Discussion}
From the comparison between the Tables~\ref{cut-flow_300_SR1} and \ref{cut-flow_300_SR2} (300 fb$^{-1}$), and also between the Tables~\ref{cut-flow_1500_SR1} and \ref{cut-flow_1500_SR2} (1500 fb$^{-1}$), we can see that the search strategy for the SR1 signal region tends to give a higher significance. However, notice that the signal region SR2 does not involve any cut in invariant mass, what makes this search strategy more model-independent than the SR1 one. Even so, the $m_{4b}$ invariant mass window considered in SR1 is broad enough to cover a large region of Higgs masses and it contains the {\it moderate mass} scenario which is in fact the scenario that we attempt to probe with the search strategies presented here. In addition, the signal region SR2 has the advantage of retaining more signal events; indeed, we expect 4 signal events for $\mathcal{L}=300\,\text{fb}^{-1}$, while just one signal event is expected when applying the SR1 strategy for the same luminosity. This increase in the signal events comes at the expense of keeping at the same time more background events and adding two more reducible backgrounds. Moreover, in SR2 the backgrounds are dominated by their reducible contribution. The presence of more background events is risky when the possible systematic uncertainties are taken into account; in fact, when such a source of uncertainties is included, the statistical significance defined in Eq.~(\ref{Sig1}) is modified as follows~\cite{Cowan:2012}:
\begin{equation}
\label{Sigsys}
{\cal S}_\text{sys} = \sqrt{2 \left((B+S) \log \left(\frac{(S+B)(B+\sigma_{B}^{2})}{B^{2}+(S+B)\sigma_{B}^{2}}\right)-\frac{B^{2}}{\sigma_{B}^{2}}\log \left(1+\frac{\sigma_{B}^{2}S}{B(B+\sigma_{B}^{2})} \right) \right)} \,,
\end{equation}
where $\sigma_{B}=(\Delta B) B$, with $\Delta B$ being the relative systematic uncertainty. From the comparison between Eqs.~(\ref{Sig1}) and (\ref{Sigsys}), it can be seen that the bigger the number of background events, the higher the degradation of the signal significance due to the systematic uncertainties.\par 
\begin{table}
\begin{center}
\begin{tabular}{r|ccrr|ccrr}
\hline \hline
&\multicolumn{4}{c|}{SR1} & \multicolumn{4}{c}{SR2} \\
${\cal L}$ [fb$^{-1}$] & Signal & Background & ${\cal S}$ & ${\cal S}_\text{sys}$ & Signal & Background & ${\cal S}$ & ${\cal S}_\text{sys}$ \\ \hline
300 & 1.07 & 0.18 & 1.62 & 1.59 & 4.19 & 5.64 & 1.59 & 1.25 \\
1000 & 3.56 & 0.62 & 2.97 & 2.80 & 13.95 & 18.80 & 2.91 & 1.67 \\
1500 & 5.34 & 0.94 & 3.63 & 3.33 & 20.93 & 28.2 & 3.56 & 1.76  \\ \hline \hline
\end{tabular}
\caption{Number of signal and background events along with the significances obtained without taking into account systematic uncertainties (see Eq.~(\ref{Sig1})) and assuming that this source of uncertainty amounts to $30\%$ (see Eq.~(\ref{Sigsys})). The results corresponding to luminosities of 300, 1000, and 1500 fb$^{-1}$ are shown.}
\label{lums}
\end{center}
\end{table}   
In Table~\ref{lums} we summarize the results for the signal regions SR1 and SR2 including the significance ${\cal S}_\text{sys}$, obtained assuming $30\%$ of systematic uncertainties. Besides presenting the results corresponding to ${\cal L}=$ 300 and 1500 fb$^{-1}$, we also display the results for an intermediate luminosity of ${\cal L}=1000$ fb$^{-1}$. From the Table~\ref{lums}, it is clear that the systematic uncertainties have a meaningful impact on the significances corresponding to SR2, while the values of the significance are slightly reduced in the case of SR1. With an integrated luminosity of $1500~\text{fb}^{-1}$, the SR1 retains 5 signal events with a significance of more than $3\sigma$, even when a $30\%$ of systematic uncertainties is considered. The results for this luminosity in the case of SR2 are similar, except that now the significance is degraded to $\sim 1.8\sigma$ due to the effect of the systematic uncertainties.  The results for the intermediate luminosity ($1000~\text{fb}^{-1}$) are also promising, achieving evidence of the new physics signal in both signal regions. Again, note that with the conservative value chosen for the systematic uncertainties ($30\%$), the significance for SR2 decreases to $\sim 1.7$. It is important to mention here that our choice of the signal cross sections is rather conservative and if we had considered an enhancement of the final cross section, we could have obtained better significances even for the lowest luminosity of 300 fb$^{-1}$. For example, an increase of a factor 2 in BR($H \to hh$)$\times$BR($A \to \chi\chi$), which is not harebrained\footnote{This enhancement is not difficult to achieve within the low-$\tan\beta$ regime of the MSSM scenarios considered here. BR($H \to hh$) can reach values up to 65\%~\cite{Bagnaschi:2015hka} and we have shown in Table~\ref{BRs} that BR($A \to \chi\chi$) may be of the same order. These two values compared with the considered ones in the {\it moderate-mass} scenario would mean an increase in BR($H \to hh$)$\times$BR($A \to \chi\chi$) of a factor $\sim$ 2, which would be directly translated to an identical growth in the total cross section of the signal.}, would mean a proportional enhancement on the signal events, resulting in statistical significances of 2.73 and 2.96 at 300 fb$^{-1}$ for SR1 and SR2, respectively. These results would not be so affected by the systematic uncertainties, with values for ${\cal S}_\text{sys}$ of 2.65 and 2.27, respectively, which are also very close to the evidence threshold.

Finally, although we have not generated enough number of background events to provide a prediction of the significances at 3000 fb$^{-1}$, we have naively assumed that the background rates scale as the cross section of the signal~\cite{Djouadi:2015jea}, which is a conservative standpoint. Under this assumption, we obtain that the statistical significances ${\cal S}$ for SR1 and SR2 at 3000 fb$^{-1}$ are 5.14 and 5.04, respectively, which reach the discovery level. On the other hand, if we assume conservatively that the 30\% systematic uncertainties will persist at these luminosities, the significances are reduced to 4.40 for SR1 and 1.89 for SR2. It is clear that, even under these conservative assumptions, the search strategies developed along this work offer a chance to show a first hint at the HL-LHC of this class of new physics signals of heavy Higgs bosons decaying invisibly.

\section{Conclusions}
\label{conclusions}

In this work we have developed a search strategy for invisible Higgs decays through heavy Higgs-pair production at the LHC, which makes plausible its detection. Our proposed alternative to the search strategy for heavy neutral Higgs bosons focuses on the production of a pair of Higgs bosons $H+A$ through the tree-level mode $q \bar q \to Z^\ast \to HA$, and considers the elusive possibility of detecting invisible pseudoscalar decays. In order to make quantitative statements, we have worked within a particular MSSM scenario, so-called Slim SUSY, but the conclusions hold in general for any given scenario with a similar mass spectrum. The only new particles at the EW scale in these MSSM scenarios are the heavy Higgs bosons and the charginos and/or the neutralinos. This fact allows for invisible Higgs decays $H, A \to \chi \chi$ if $M_{H,A} > 2M_\chi$. Within the {\it moderate-mass} scenario ($M_A = 300$ GeV) the channel $q \bar q \to HA \to 4b+E_T^{\text{miss}}$ ($H \to hh \to 4b$ and $A \to \chi\chi$) is found to be the most promising signature to probe invisible Higgs decays through heavy Higgs-pair production at the LHC.
We have shown that the invisible channel is the dominant decay mode of the pseudoscalar both for the light and moderate mass scenarios, so that the cross sections of those channels that involve invisible particles in the final state are higher than those that lead to visible final states. This, together with the possibility of using powerful discriminator variables based on $E_T^\text{miss}$ significance, makes the search for heavy neutral scalars through their invisible decays more auspicious than considering visible final states. It is important to note that this is not the case for the large mass scenario, which is clearly dominated by the four tops channel, where both heavy scalars decay visibly (see Table~\ref{ratesHM}).

We have defined two different signal regions, namely, SR1 with exactly 4 $b$-tagged jets and SR2, in which the $b$-tagging requirement is relaxed to 3 or 4 $b$-tagged jets. A detailed search strategy has been performed in each signal region, based on a sequential optimization of the corresponding cuts that maximizes the statistical significance in each case. As a result, we have obtained prospects of evidence ($\sim$ 3$\sigma$) at 1000 fb$^{-1}$ and a conceivably possibility of discovery ($\sim$ 5$\sigma$) at 3000 fb$^{-1}$ in both of the two signal regions. If we take into account systematic uncertainties of 30\%, the evidence significance is degraded to 2.80 (1.67) for the signal region SR1 (SR2), whilst the discovery significance reduces to 4.40 (1.89). We can conclude that the search strategies developed along this work offer the opportunity to discover this class of new physics signals of heavy Higgs bosons decaying invisibly at the HL-LHC.

On the other hand, although a thoughtful analysis of the reducible backgrounds is beyond the scope of this work, our results are conservative. In principle, it should not be difficult to improve them. A better cut optimization through multivariate analysis (MVA) with the boosted decision tree (BDT) algorithm might ameliorate the signal/background ratio. Our estimation of the systematic uncertainties is very rough and one could expect a significant reduction in the future. In addition, although a combination of significances of both signal regions (SR1 \& SR2) would imply dealing with large correlations and thus making it highly nontrivial, it might result in an increase of the total significance, with potential interest even for the lowest luminosity of 300 fb$^{-1}$ considered here. Furthermore, if it were possible to keep the same value of the signal cross section for larger values of the heavy Higgs bosons masses (increasing, for instance, the rates of the decay channels $H \to hh$ and $A \to \chi\chi$), the $E_T^\text{miss}$ in the signal events would be greater than in the case of the {\it moderate-mass} scenario. Therefore, a more stringent cut on the $E_T^\text{miss}$ significance would lead to an improvement on the significances. Last but not least, larger signal cross sections could be generated within the MSSM scenarios considered in this work or in other BSM models that drive us to the same final state topology. For example, it is not harebrained to obtain an increase of a factor of 2 in BR($H \to hh$)$\times$BR($A \to \chi\chi$), which would mean a proportional enhancement on the signal events, resulting in statistical significances of 2.73 and 2.96 at 300 fb$^{-1}$ for SR1 and SR2, respectively. Considering systematic uncertainties of 30\%, these significances would be 2.65 and 2.27, respectively, which are also close to the evidence level.

\section*{Acknowledgments}

The authors thank Aurelio Juste, An\'{\i}bal Medina, Hern\'an Wahlberg, and Jos\'e Zurita for fruitful discussions.
This work has been partially supported by CONICET and ANPCyT project PICT-2013-2266 (E.A., N.M., R.M., and A.S.).
J.L.D.C. acknowledges support from VIEP-BUAP and CONACYT-SNI (Mexico).

%\newpage
% \baselineskip 16pt
\bibliographystyle{unsrt}

\end{document}